\documentclass{emulateapj}

\slugcomment{Submitted to ApJ March 7, 2015; accepted April 2, 2015}

\shorttitle{PKS 0637-752 and TeV Emission from Quasar Jets}
\shortauthors{Meyer et al.}

\begin{document}
\newcommand{\btxt}[1]{{\bf #1}}

\title{Ruling out IC/CMB X-rays in PKS 0637-752 and the Implications
  for TeV Emission from Large-Scale Quasar Jets}


\author{Eileen T. Meyer\altaffilmark{1}}
\affil{Space Telescope Science Institute, Baltimore, MD 21218, USA}
\email{meyer@stsci.edu}

\author{Markos Georganopoulos\altaffilmark{2}}
\affil{University of Maryland Baltimore County, Baltimore, MD 21250, USA}

\author{William B. Sparks}
\affil{Space Telescope Science Institute, Baltimore, MD 21218, USA}

\author{Leith Godfrey}
\affil{ASTRON, Netherlands Institute for Radio Astronomy, P.O. Box 2, 7990 AA Dwingeloo, Netherlands}

\author{James E. J. Lovell}
\affil{School of Physical Sciences, University of Tasmania, Private Bag 37, Hobart, Tasmania 7001, Australia}

\and

\author{Eric Perlman}
\affil{Florida Institute of Technology, Melbourne, FL 32901, USA}

\altaffiltext{1}{University of Maryland Baltimore County, Baltimore, MD 21250, USA}
\altaffiltext{2}{NASA Goddard Space Flight Center, Greenbelt, MD 20771, USA}



\begin{abstract}
The \emph{Chandra} X-ray observatory has discovered dozens of
resolved, kiloparsec-scale jets associated with powerful quasars in
which the X-ray fluxes are observed to be much higher than the
expected level based on the radio-optical synchrotron spectrum. The
most popular explanation for the anomalously high and hard X-ray
fluxes is that these jets do not decelerate significantly by the
kiloparsec scale, but rather remain highly relativistic (Lorentz
factors $\Gamma\sim$10). By adopting a small angle to the
line-of-sight, the X-rays can thus be explained by inverse Compton
upscattering of CMB photons (IC/CMB), where the observed emission is
strongly Doppler boosted. Using over six years of \emph{Fermi}
monitoring data, we show that the expected hard, steady gamma-ray
emission implied by the IC/CMB model is not seen in PKS 0637-752, the
prototype jet for which this model was first proposed. IC/CMB emission
is thus ruled out as the source of the X-rays, joining recent results
for the jets in 3C 273 \citep[using the same method;][]{mey14} and PKS
1136-135 \citep[using UV polarization;][]{cara2013}. We further show
that the \emph{Fermi} observations give an upper limit of $\delta<$6.5
for the four brightest X-ray knots of PKS~0637-752, and derive an updated
limit of $\delta<$7.8 for knots A and B1 of 3C~273 (assuming
equipartition). Finally, we discuss the fact that high levels of
synchrotron X-ray emission in a slow jet will unavoidably lead to a
level of angle-integrated TeV emission which exceeds that of the TeV
BL Lac class.
\end{abstract}





\section{Introduction}

In August 1999, the \emph{Chandra} X-ray Observatory observed its
first celestial target, quasar PKS~0637-752, during the initial
focusing of the telescope \citep{schwartz2000,chartas2000}. Along with
the bright quasar core, \emph{Chandra} unexpectedly detected X-rays
from the kiloparsec (kpc) scale relativistic jet (previously known
from radio imaging, Figure~1). Unlike the synchrotron spectrum of
lower-power FR I jets like M87 which easily extend up to X-ray
energies \citep[e.g.][]{wilson2002}, the synchrotron spectrum of
powerful kpc-scale jets (including PKS~0637-752) generally peak at or
below the IR/Optical band. The X-rays detected in the kpc-scale jet of
PKS~0637-752 were much brighter than expected from the radio-optical
synchrotron spectrum, or indeed from either synchrotron self-Compton
(SSC) or inverse Compton upscattering of ambient CMB photons (IC/CMB)
assuming mildly relativistic kpc-scale flows under equipartition
conditions \citep{chartas2000}.  Further, the X-ray spectrum of the jet
was remarkably hard, with a photon index of $1.76\pm0.1$
(Figure~\ref{fig:schematic}).

Multi-epoch measurements of sub-parsec scale jets of powerful quasars
with Very Long Baseline Interferometry (VLBI) have detected
superluminal proper motions which imply that these jets start out
highly relativistic, with Lorentz factors ($\Gamma$) of 10-50
\citep{jor05,lis09}. Though it had long been supposed based on
population studies that jets decelerate and are at most mildly
relativistic by the time they reach the kpc scale
\citep[e.g.][]{arshakian2004,mullin2009}, no direct measurements have
confirmed this. \cite{tav00} and \cite{cel01} thus suggested that the
X-rays from the jet in PKS~0637-752 could be explained by IC/CMB
emission if the jet \emph{remained} highly relativistic
($\Gamma\sim$10), and was pointed at a fairly small angle to our line
of sight (6$^\circ$). This produces a much larger Doppler boosting
factor ($\delta\sim$10) and enables the IC/CMB X-ray flux to match the
observations.

Over the past decade and a half since the launch of \emph{Chandra},
dozens more kpc-scale quasar jets with anomalously hard and/or high
X-ray fluxes have been detected
\citep[e.g.][]{sambruna2001_3c273,sambruna2002,aneta2003,sambruna2004,marshall2005,harris2006,siemiginowska2007,marshall2011,kharb2012,godfrey2012_2101}. The
IC/CMB model has been by far the most popular explanation of these
X-rays, though problems have been noted \citep[e.g][]{har06,harris2006}.  Besides the
unconfirmed fast speeds required on the kpc scale, IC/CMB often
requires the jet to be pointed very close along our line-of-sight,
leading to a deprojected jet length longer than 1 Mpc, comparable to,
or greater than the largest known giant radio galaxies
\citep{der04}. Further, the electrons responsible for upscattering the
CMB into the \emph{Chandra} band are at much lower energies than are
traced by radio observations. This extension of the electron energy
distribution is energetically costly, in some cases leading to
`super-Eddington' jet power requirements
\citep{der04,uch06}. \cite{geo04} also noted that the commonly
observed severe decrease in X-ray to radio flux ratio with distance
down the jet \citep[e.g. 3C~273,][]{jes06} can only be reconciled with
IC/CMB X-rays if the jet is strongly decelerating. \cite{har06},
seeking to verify this for a sample of quasar jets, also noted that
this explanation still leaves significant discrepancies with
observations, namely the lack of similar deceleration profiles at
kpc-scales for more misaligned jets. These problems lead to the
suggestion that the X-rays could alternatively be synchrotron emission
from a second electron population in the jet, albeit of unknown origin
\citep{atoyan2004,harris2004,kataoka2005,har06,jes06,uch06}.

Despite a significant effort by the community to amass high-resolution
radio, optical, and X-ray imaging of dozens of quasar jets, the
fundamental problem up to now has been that fitting the radio-to-X-ray
spectral energy distribution (SED) alone cannot distinguish between
the IC/CMB and synchrotron explanations for the X-rays
\citep{cara2013}. The difference in power requirements between the two
mechanisms is great, as is the extremely different idea of jet
structure that they imply. Discriminating between these models is
essential to make progress on the the physics of jets and their impact
on their environments.

\begin{figure}[!ht]
\begin{center}
\includegraphics[width=3in]{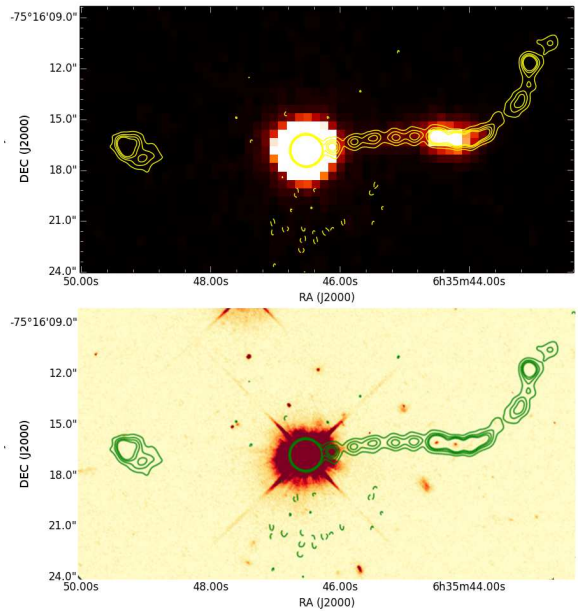}
\caption{\emph{Upper Panel:} \emph{Chandra} X-ray image of
  PKS~0637-752 \citep{chartas2000}, with ATCA 18 GHz radio contours
  overlaid \citep{godfrey2012_pks0637}. \emph{Lower Panel:} Optical
  image of PKS~0637-752, taken with ACS/WFC on HST in F475W
  \citep{mehta2009} with ATCA 18 GHz radio contours overlaid.}
\end{center}
\label{fig:pks0637_panels}
\end{figure}

It was with an eye to resolving this long-standing deadlock that
\citet{geo06}, hereafter {\bf G06}, suggested that \emph{Fermi}
Large Area Telescope (LAT) observations could confirm or rule out the
IC/CMB mechanism for the X-rays, by detecting (or not) the high level
of gamma-ray emission this mechanism predicts
(Figure~\ref{fig:schematic}). We have previously looked for this
gamma-ray emission from the jet of 3C~273, and ruled out IC/CMB
gamma-rays from (the brightest) knot A alone at the $>$95\% level, and
from knots A through D1 combined at the $>$99.9\% level
\citep[][hereafter {\bf M14}]{mey14}.

In this paper we report new \emph{Fermi} observations of PKS~0637-752
which show that the expected steady gamma-ray emission from the IC/CMB
mechanism is also ruled out by deep upper limits at the 99.9\%
level. We also present updated limits for 3C ~273, showing that the
expected gamma-rays from IC/CMB are now ruled out at the 99.99\% level
in more than one \emph{Fermi} energy band. We show that the deep upper
limits at GeV energies place interesting constraints on the Doppler
beaming factors which implies that these jets are not highly
relativistic on the kpc scale.  We further discuss the surprising
implications of slow, synchrotron X-ray jets on our understanding of
the total radiative output of quasars, especially at TeV energies.

\section{Methods}
\label{sec:methods}

\subsection{The \emph{Fermi} Test of IC/CMB}

\begin{figure}[!t]
\begin{center}
\includegraphics[width=3.5in]{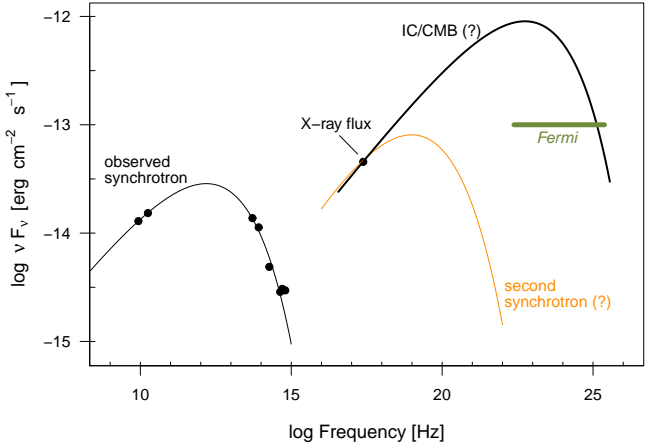}
\caption{
\label{fig:schematic}A depiction of the \emph{Fermi} test for the X-ray emission
  mechanism in large-scale jets. Here we show the radio, optical, and
  X-ray photometry for the knots (total) in the jet of PKS~0637-752 as
  black points.  The thin black line shows the synchrotron fit to the
  data. This synchrotron spectrum must be shifted by a factor
  3.4$\times 10^{10}$ in frequency and 3.05$\times 10^2$ in luminosity,
  corresponding to $B/\delta$ = 1.5$\times 10^{-7}$ G, in order for
  the IC/CMB spectrum (thick black line) to match the X-ray flux
  level. As shown, this implies a high level of gamma-ray emission
  which should be detectable with \emph{Fermi}. Alternatively, the
  X-rays could be due to a second synchrotron component (thin orange
  line).}
\end{center}
\end{figure}

As first noted by G06, the shape of the IC/CMB spectrum is constrained
to match the synchrotron spectrum, with a shift in frequency and
luminosity solely determined by the factor $B/\delta$ where $\delta$
is the Doppler beaming factor and $B$ the magnetic field
strength. From G06, we have:

\begin{equation}
\frac{\nu_c}{\nu_s} = \frac{\nu_\mathrm{CMB}\delta^2\gamma^2}{eB\delta\gamma^2/[2\pi m_e c(1+z)]}
\end{equation}

\begin{equation}
\frac{L_c}{L_s} = \frac{32\pi U_\mathrm{CMB}(1+z)^4\delta^4}{3(B\delta)^2}, 
\end{equation}
where $\nu_c$ and $\nu_s$ ($L_c$, $L_s$) are the observed Compton and
synchrotron frequencies (luminosities) emitted by electrons of Lorentz
factor $\gamma$, $e$ and $m_e$ are the electron charge and mass, and
$\nu_\mathrm{CMB}$ = 1.6$\times 10^{11}$ Hz is the CMB peak frequency
at z = 0. However, if the observed X-ray fluxes are to be produced by
the IC/CMB mechanism, then the value of $B/\delta$ is already uniquely
determined by the requirement to match the X-ray flux level, at which
point there is no freedom at all in the rest of the spectrum.  The
peak of the IC/CMB spectrum will fall in the GeV band, as shown in
Figure~\ref{fig:schematic}. Note that this prediction is not
predicated on any particular (e.g., equipartition) magnetic field
strength.

The \emph{Fermi}/LAT lacks the spatial resolution to detect the jet
separately from the gamma-ray bright quasar core, which is only 10$''$
away -- the \emph{Fermi}/LAT 68\% containment radius is on the order
of tenths of a degree to degrees. However, in powerful quasars the
inverse Compton core emission generally peaks at a few MeV, producing
a soft and extremely variable spectrum in the \emph{Fermi} band, with
long periods of relative quiescence. Indeed, PKS~0637-752 was detected
in the 2nd \emph{Fermi} source catalog \citep[2FGL,][]{nolan2012} as
source 2FGL~J0635.5-7516, with a very soft spectrum (photon index of
$\Gamma_p$=2.71) and high degree of variability (variability index =
347). In contrast, the IC/CMB emission from the large scale jet is
expected to be harder and completely non-variable. The latter property
allows us to combine the \emph{Fermi} data taken only when the quasar
core is in a low state to try to detect or place limits on the IC/CMB
emission.

\vspace{15pt}
\subsection{\emph{Fermi} Analysis of PKS 0637-752}

We first combined the all-sky weekly LAT event and spacecraft files
for weeks 9 through 325 of the \emph{Fermi} mission, corresponding to
\emph{Fermi} Mission Elapsed Time (MET) from 239557417 to 430608212 and
calendar dates 2008 August 4 to 2014 August 24. In order to analyze
the region around PKS~0637-752, we used the publicly available
`quickAnalysis' script\footnote{The public scripts mentioned in the
  text are available at
  http://fermi.gsfc.nasa.gov/ssc/data/analysis/user/.} to run the
\emph{Fermi} analysis tools and generate the filtered event file,
livetime cube, and exposure map, using a region of interest (ROI) of
10$^\circ$ and an otherwise default configuration. The starting source
list was generated from the publicly available {\tt make2FGLxml}
script, which generates the xml file pre-populated with 2FGL catalog
sources. We used a binned maximum likelihood to get an initial fit for
all the catalog sources in our ROI. We also included sources a further
5$^\circ$ out from our ROI, but with spectral parameters fixed to the
2FGL catalog values. PKS~0637-752 was detected with a very high
test-statistic (TS, roughly significance squared) value of 289, a
100~MeV to 100~GeV photon flux of 3.16$\times 10^{-8}$ s$^{-1}$
cm$^{-2}$, and a photon index $\Gamma_p$ = 2.64, similar to the value
reported in the 2FGL catalog.

\begin{figure}[!b]
\begin{center}
\includegraphics[width=3.5in]{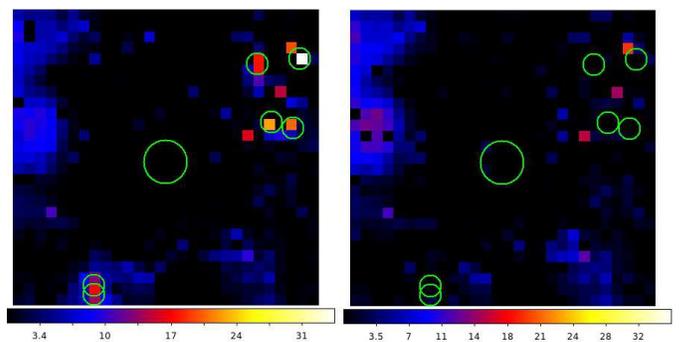}
\caption{
\label{fig:initTSmap}\emph{Left: }An initial TS map of the region around
  PKS~0637-752, showing the excess TS present in 0.5$^\circ$ pixels
  over the best-fit likelihood model using 2FGL catalog sources.  The
  large circle marks the position of PKS~0637$-$752. The smaller
  circles mark the positions of the six new sources not present in the
  2FGL catalog (corresponding to regions of excess TS with pixel
  values $>$20). \emph{Right: } The updated TS map (same field-of-view and
  binning) after the six new sources were localized and fit with a binned
  likelihood.}
\end{center}
\end{figure}

We checked for additional significant sources within 7$^\circ$ of
PKS~0637-752 but not in the 2-year LAT catalog by making a TS residual
map. The TS map was created by first freezing all source parameters in
the initial model to their best-fit values, and then checking what TS
value a test point source at different locations would be given. The
resulting initial TS map is shown at left in
Figure~\ref{fig:initTSmap}. In two areas the coarse 0.5$^\circ$
binning did not allow separation of what appeared to be multiple
components, and so a finer 0.125$^\circ$ binning was used over these
smaller areas to determine the rough location of new sources.

\begin{deluxetable}{ccc|ccc}
  \tabletypesize{\footnote}
  \tablecolumns{6}
  \tablewidth{0pt}
  \tablecaption{\label{table:newsources} New sources in ROI of Targets}
  \tablehead{
    \multicolumn{3}{c}{\underline{near PKS 0637-752}} & \multicolumn{3}{c}{\underline{near 3C 273}} \\
    RA$_{J2000}$  & Dec$_{J2000}$ & TS & RA$_{J2000}$   & Dec$_{J2000}$  & TS  \\
    (deg) & (deg) &  & (deg) & (deg) &
  }
\startdata  
\phantom{0}82.43046 & -72.74572 &           57.5 & 192.82676 &           -2.00263 & \phantom{0}70.7 \\
\phantom{0}81.13040 & -69.60575 &           60.5 & 190.96827 &           -2.29561 & \phantom{0}82.3 \\
\phantom{0}78.98804 & -72.72570 &           28.6 & 187.15935 &           -3.29476 & \phantom{0}60.2 \\
\phantom{0}86.33797 & -70.34640 &           39.6 & 184.47577 &           -0.48239 & \phantom{0}92.5 \\
          119.70430 & -80.70430 & \phantom{0}9.9 & 193.43690 & \phantom{-}3.47391 &           224.6 \\
          118.83300 & -80.32970 &           16.1 & 192.63313 & \phantom{-}2.25116 & \phantom{0}65.7
\enddata
\end{deluxetable}

Rough starting positions of apparent new significant sources were
measured from the TS map by hand, only considering as candidates those
with a central pixel value (TS) $>$ 20. Each potential new source was
then localized one at a time as follows. A power-law point source was
added at the rough location of the excess TS, with normalization and
powerlaw index free. Sources within 5$^\circ$ of the new source also
had normalization/index parameters free. We ran a binned likelihood to
get a starting fit for the new source plus other sources in the area.
The fit parameters were then frozen again, so that we could next
optimize the RA/Dec location of the new source.  The existing tool,
{\tt gtfindsrc}, only works for unbinned likelihood analysis, so we
built our own binned version of the tool which works in the same way.
Using the frozen model, we used the python {\tt minimize} function in
the {\tt scipy} package (L-BFGS-B method) to optimize the
log-likelihood value versus the RA and Dec position, given a
reasonable range of about 1 degree around the starting positions noted
by hand. We then update our new source in the model with this
optimized position (positions always remain fixed when optimizing over
spectral parameters). We un-freeze the normalization and index of
the new source and the local sources and run a final binned likelihood
optimization to get an updated model of all sources in the ROI.  This
is then the starting model for the next new source to be added, until
all new sources have been localized and a final optimized fit
derived. For the 7$^\circ$ ROI around PKS~0637-752, six new sources
were added to the model, and an updated TS map run from this larger
source list shows that the excess TS previously seen is now gone
(right panel of Figure~\ref{fig:initTSmap}). A list of the new sources
with their location and TS value is given at left in
Table~\ref{table:newsources}.

\begin{figure}[!b]
\begin{center}
\includegraphics[width=3.4in]{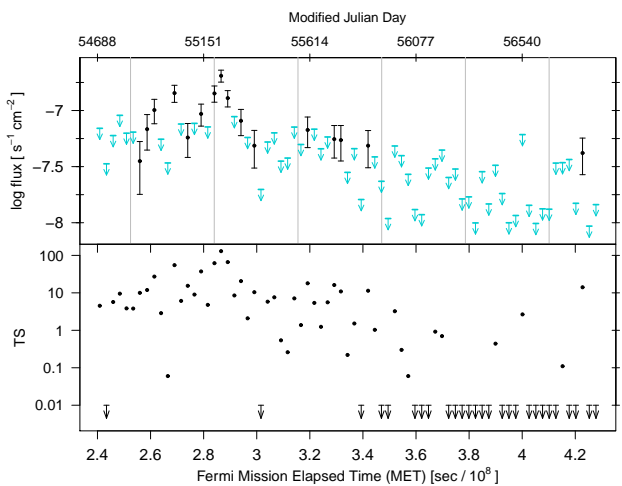}
\caption{\label{fig:0637_lightcurve}\emph{Upper: }Lightcurve of
  PKS~0637-752. The total 100 MeV - 100 GeV photon flux for
  PKS~0637-752 in 10.5-day (total GTI) bins versus the mean Mission
  Elapsed Time (MET) of the bin. Upper limits are shown where
  TS$<$10. The gray vertical lines correspond to the start of a
  calendar year, beginning with 2009. \emph{Lower: } The TS value
  of the bin versus MET. }
\end{center}

\end{figure}
As the PKS~0637-752 core is a significant \emph{Fermi} source, our
approach to detecting and/or setting limits on the IC/CMB gamma-ray
emission exploits the variability of the core which cannot be
spatially resolved separately from the large-scale jet due to the poor
angular resolution of \emph{Fermi}. During times when the blazar is
quiescent, the hard, steady emission from IC/CMB will either appear as
a steady plateau, or else the upper limits generated will place
constraints on the level of the IC/CMB emission.  

In order to build a lightcurve of the core, the full 6-year dataset
was divided into bins of equal good time interval (GTI) time, totaling
10.5 days.  We then used our updated (2FGL + 6 new) model described
above and ran a binned likelihood to fit PKS~0637-752 as a power-law
source, with sources more than 5$^\circ$ away fixed.  The resulting
lightcurve for the core over the full time range is shown in
Figure~\ref{fig:0637_lightcurve}, with the 100 MeV - 100 GeV photon
flux shown on top and the corresponding TS shown below. The clear
variability in the light curve indicates that the total \emph{Fermi}
flux is dominated by the core.

\begin{figure}[!b]
\begin{center}
\includegraphics[width=3.4in]{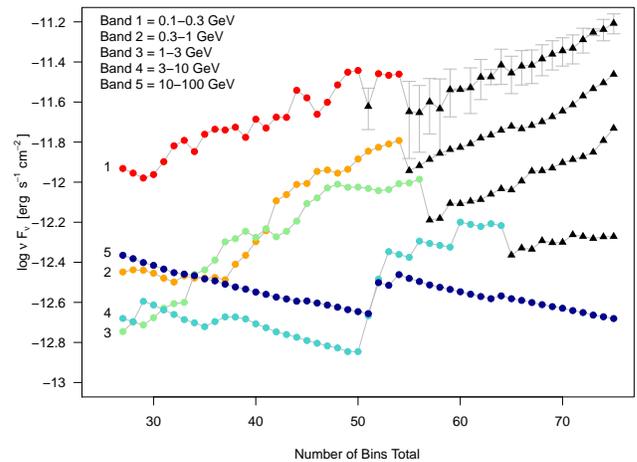}
\caption{\label{fig:pks0637_limits}The results of the progressive binning
  analysis on PKS~0637-752. The upper axis gives the $\nu F_\nu$ flux while lower axis
  gives the total number of bins combined (where ordering is based on
  TS value and not time order), starting from the 27 bins with
  TS$<$0.01. The points give the upper limits (colored dots) or
  detected fluxes (TS$>$10, black triangles) for each of the five
  \emph{Fermi} energy bands (red, orange, green, cyan, navy from
  lowest to highest energy). Errors on the detected fluxes are only
  shown for the lowest-energy band for clarity but are similar across
  bands. For the highest-energy band (10 - 100 GeV) no detection is ever
  made. The increasing detected fluxes indicate that the quasar core is being
  detected in the other bands.}
\end{center}
\end{figure}

\begin{figure*}[t]
\begin{center} 
\includegraphics[width=6in]{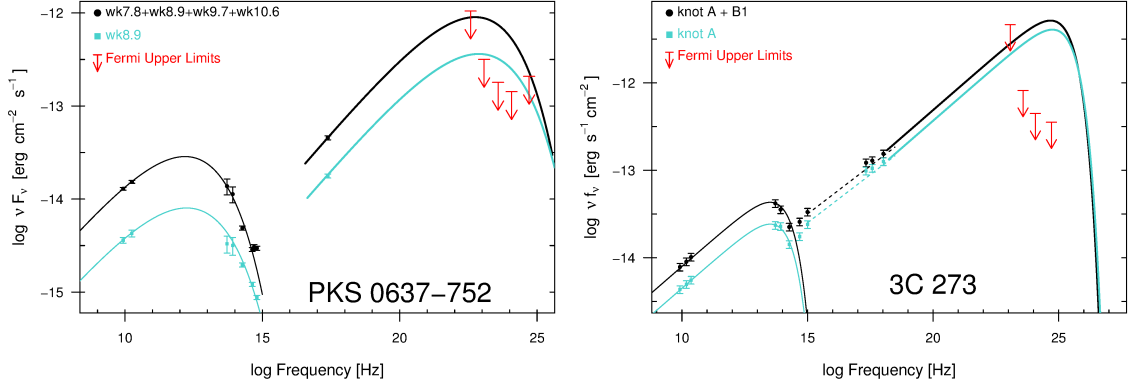}
\caption{\label{fig:both_results}\emph{Left: }The SED for the
  large-scale jet of PKS~0637-752.  Data for the four brightest X-ray
  detected knots combined is shown as black points. The SED of the
  X-ray brightest knot, wk8.9, is plotted with blue points.
  \emph{Right: }The SED for the knots of 3C~273, with black points for
  knots A and B1 combined, and blue for X-ray brightest knot A only. For both plots, the
  \emph{Fermi} 95\% upper limits are shown in red.
}
\end{center}
\end{figure*}

We next began a `progressive binning' analysis, in which the
lightcurve bins were ordered from lowest to highest TS value. Of the
entire set of 75 time bins, 27 showed a TS level consistent with zero
for the location of PKS 0637-752 (upper limits in lower panel of
Figure~\ref{fig:0637_lightcurve}).  Starting from these 27 bins
combined, we progressively combine the event files for the lowest bins
plus the next lowest bin in TS, at each step optimizing the fit of
PKS~0637-752 and the sources within 5$^\circ$ with a binned
likelihood. We repeated adding the next-highest bin and getting the
maximum likelihood fit until all bins had been added together. Note
that this re-combining of the lightcurve in a discontinuous way is
appropriate for deriving a limit on the large-scale jet because the
IC/CMB emission is predicted to be completely non-variable, and thus
there is no risk of any selection effect via variability. The variable
core clearly dominates the flux levels determining the ordering, and
is disconnected from the jet in any case.  At each step we evaluated
the TS and flux level in the five canonical \emph{Fermi} energy bands
of 0.1-0.3, 0.3-1, 1-3, 3-10 and 10-100 GeV, calculating the 95\%
upper limit flux value when TS$<$10 in any given bin. Previous work on
3C~273 has shown that the exact method of ordering the bins (whether
by using the TS value or the total flux or upper limit value for the
bin) does not significantly affect the order or resulting upper limits
(M14).

As shown in Figure~\ref{fig:pks0637_limits}, the highest energy bands
gave upper limits which decreased with the increasing exposure as more
bins are added, where we have color-coded the upper limits in the 5
energy bands, and black triangles indicate a significant detection in
the band. 
Note that in the lower-energy bins, where the PKS~0637-752 quasar core
dominates due to its soft spectrum, the upper limits reach a minimum
rather quickly, and generally increase before becoming detections. It
must be noted that the detected gamma-ray emission in these bands is from
the quasar core, not the large-scale jet, based on the soft spectrum,
and the fact that the emission level rises as more bins are added
(showing that the source is indeed variable and that the bins are
ordered by flux level). If we were to detect the large-scale jet, we
would expect it to contribute significantly in the two highest-energy
\emph{Fermi} bands, and also to remain steady in flux level after the
exposure became long enough to detect it. This latter point is
important for distinguishing a detection of IC/CMB gamma-rays versus
the tail of the Compton emission from the core, which should (like the
low-energy bins) rise in flux as more bins are added. While the fourth
energy band detected points do not rise as quickly as the first three,
the flux level is far above the upper limit derived after 50 bins, so
cannot be the steady emission of the large-scale jet, which must be
below this limit. The highest bin never shows a significant detection
of either component.

\vspace{15pt}
\subsection{Fermi Analysis of 3C 273}

We re-analyzed the \emph{Fermi} data for 3C 273 using the 6-year
dataset to compare with the results from M14 using 4.5 years of data,
as the core remained relatively quiescent over the additional time
elapsed. We followed the same procedure as outlined above for
PKS~0637-752, finding six new sources within 7$^\circ$ of the position
of 3C~273, listed at right in Table~\ref{table:newsources}. The core of 3C~273
was detected with a TS of 17504, with a 100 MeV - 100 GeV photon flux
of 3.68$\times 10^{-7}$ s$^{-1}$ cm$^{-2}$ and $\Gamma_p$ = 2.67. A
lightcurve was made using bins totaling 10.5 days in GTI time, and
ordered according to TS (a total of 88 bins). The progressive-binning
was started from the single lowest bin, with the next-highest bin
continually added as described above until all bins were added. At
each step the flux (or 95\% upper limit) was calculated for the five
canonical \emph{Fermi} energy bins, as for PKS~0637-752.

\begin{deluxetable*}{lccccccclcl}[t]
\tabletypesize{\scriptsize}
\tablecaption{Results of the Fermi Data Analysis}
\centering
\tablehead{
Source & Band & $E_1$ & $E_2$ & log Freq. & 95\% Limit              & Bins & \multicolumn{2}{c}{Combined Knots\tablenotemark{*}} & \multicolumn{2}{c}{Single Knot\tablenotemark{$\dagger$}} \\
       &      & (GeV) & (GeV) &   (Hz)    & (erg s$^{-1}$ cm$^{-2}$)  & Added &  $F_\mathrm{IC/CMB}$ & \% Ruled & $F_\mathrm{IC/CMB}$ & \% Ruled\\
       &      &       &       &           &                         & & (erg s$^{-1}$ cm$^{-2}$) & Out  & (erg s$^{-1}$ cm$^{-2}$) &  Out    \\
(1)    & (2)  & (3)   & (4)   & (5)       & (6) & (7) & (8) & (9) & (10) & (11) \\
}
\startdata
0637-752 & 1 & \phantom{00}0.1           & \phantom{00}{0.3}               & 22.6 & 1.05$\times 10^{-12}$     & 29     & 9.0$\times 10^{-13}$ & 92.9  & 3.6$\times 10^{-13}$ & ...  \\
             & 2 & \phantom{00}0.3           & \phantom{00}{1}\phantom{00} & 23.1 & 3.17$\times 10^{-13}$ & 32 & 8.8$\times 10^{-13}$ & 99.8  & 3.6$\times 10^{-13}$ & 94.5 \\
             & 3 & \phantom{00}1\phantom{00} & \phantom{00}{3}\phantom{00} & 23.6 & 1.80$\times 10^{-13}$ & 27 & 7.4$\times 10^{-13}$ & 99.98 & 3.2$\times 10^{-13}$ & 98.7 \\
             & 4 & \phantom{00}3\phantom{00} & \phantom{0}{10}\phantom{00} & 24.1 & 1.43$\times 10^{-13}$ & 50   & 5.3$\times 10^{-13}$ & 99.95 & 2.5$\times 10^{-13}$ & 98.6 \\
             & 5 & \phantom{0}10\phantom{00} &            100\phantom{00}  & 24.7 & 2.09$\times 10^{-13}$ & 75   & 2.3$\times 10^{-13}$ & 95.9  & 1.3$\times 10^{-13}$ & ...  \\
\\
3C 273       & 1 & \phantom{00}0.1           & \phantom{00}{0.3}           & 22.6 & 2.72$\times 10^{-11}$ & \phantom{0}1 & 2.1$\times 10^{-12}$ & ...      & 1.6$\times 10^{-12}$ & ...      \\
             & 2 & \phantom{00}0.3           & \phantom{00}{1}\phantom{00} & 23.1 & 4.63$\times 10^{-12}$ & \phantom{0}2 & 2.8$\times 10^{-12}$ & ...      & 2.1$\times 10^{-12}$ & ...      \\
             & 3 & \phantom{00}1\phantom{00} & \phantom{00}{3}\phantom{00} & 23.6 & 8.20$\times 10^{-13}$ & \phantom{0}5 & 3.6$\times 10^{-12}$ & $>$99.99 & 2.8$\times 10^{-12}$ & $>$99.99 \\
             & 4 & \phantom{00}3\phantom{00} & \phantom{0}{10}\phantom{00} & 24.1 & 4.46$\times 10^{-13}$ &          31  & 4.5$\times 10^{-12}$ & $>$99.99 & 3.5$\times 10^{-12}$ & $>$99.99 \\
             & 5 & \phantom{0}10\phantom{00} &            100\phantom{00}  & 24.7 & 3.56$\times 10^{-13}$ &          30  & 5.2$\times 10^{-12}$ & $>$99.99 & 4.1$\times 10^{-12}$ & $>$99.99 
\enddata
\label{table:limits} 

\tablenotetext{*}{Combined Knots are wk7.8, wk8.9, wk9.7, and wk10.6 for PKS~0637-752, knots A and B1 for 3C~273}
\tablenotetext{$\dagger$}{Single Knots are wk8.9 for PKS~0637-752 and knot A for 3C 273}
\end{deluxetable*}

\section{Results: Testing the IC/CMB Model}
\label{sec:results}

We show in Figure~\ref{fig:both_results} the radio to X-ray SEDs for
the jet of both PKS~0637-752 (left) and 3C~273 (right). For
PKS~0637-752 we have taken both the \emph{Chandra} X-ray and Hubble
Space Telescope (HST) infrared and optical data (NICMOS, WFPC2 and
ACS) from \cite{mehta2009}. We have also re-derived the \emph{Spitzer}
infrared fluxes for the brightest complex of knots (wk7.8, wk8.9,
wk9.7, wk10.7), following the same methods reported in
\cite{uchiyama2005}. 
We have also measured updated radio fluxes based on a re-analysis of
archival and new ATCA data at 4.8, 8.4, and 17.8 GHz \citep[first
  published in][]{godfrey2012_pks0637}. For 3C~273, data is taken from
\cite{jester2005,jes06} and \cite{uch06} and references therein.

In both figures, we consider two scenarios: the first combines the
photometry of the brightest/nearest knots to the core in order to test
the IC/CMB prediction (black points and lines).  In 3C~273,
\cite{jester2005} have already shown that only knots A and B1 have
X-ray indices similar to their radio indices (which is required for
IC/CMB), so our ``combined knot'' scenario includes only these two
knots. For PKS~0637-752, we use all the bright X-ray knots just before
the turn in the jet (wk7.8, wk8.9, wk9.7, and wk10.6) where one might
assume that some deceleration likely takes place. We do not include
the only other X-ray detected knot (wk5.7) because it is not
consistently detected at other wavelengths. The second scenario
assumes that the X-rays from the weaker knots are already \emph{not}
from IC/CMB, and so only the photometry of the X-ray brightest knot is
plotted: knot A in 3C~273 and wk8.9 in PKS~0637-752, in both cases
plotted as blue points and lines. The thin solid lines through
radio-optical points show a (phenomenological) synchrotron spectrum
fitting the data, while the heavy line shows the corresponding IC/CMB
curve to match the X-ray flux levels. As shown, for both jets, the
95\% upper limits in several bands violate the IC/CMB predictions
under either scenario.



We report in Table~\ref{table:limits} a summary of the \emph{Fermi}
data analysis for PKS~0637-752 and 3C~273. We list the definition of
the energy bins in columns 2-5, followed by the deepest 95\% upper
limit flux level (in $\nu F_\nu$) reached in our progressive binning
for each energy bin in column 6. The corresponding number of bins
co-added to reach that limit is given in column 7.  In column 8 we
list the flux predicted under the IC/CMB at the frequency given in
column 5. This flux corresponds to the IC/CMB model prediction for the
combination of knots wk7.8, wk8.9, wk9.7, and wk10.7 in PKS~0637-752,
and knots A and B1 for 3C~273.  In column 10, we have calculated at
what significance level our observations rule out the level of
predicted IC/CMB flux given in column 9. For the final two columns, we
also give the predicted flux under IC/CMB and the significance-level
that we can rule it out, but only for the X-ray brightest knot of each
jet (wk8.9 and knot A, respectively).  As shown, the IC/CMB model is
ruled out at a $>$ 99.9\% level for PKS~0637-752 and at $>$ 99.99\%
level for 3C~273 under the combined knot scenarios, and ruled out at
the 98.7\% level for PKS~0637-752 and at $>$ 99.99\% level for 3C~273
when considering only the single brightest knot.

\section{Discussion}
\label{sec:discussion}

These two cases where the IC/CMB origin for the X-rays has been
unambiguously ruled out join with that of PKS~1136-135, where high UV
polarization has shown that the second component (UV to X-ray) must be
synchrotron in origin, since significant polarization is not expected
in the IC/CMB scenario \citep{cara2013}. The UV polarization method is
unfortunately not able to be applied in general, as not all of the known
quasar jets show the second component already dominating in the UV,
being instead dominated by the radio-optical synchrotron component,
though we note that 3C~273 could be confirmed in this way.  

\begin{figure*}[!t]
\begin{center}
\includegraphics[width=6.5in]{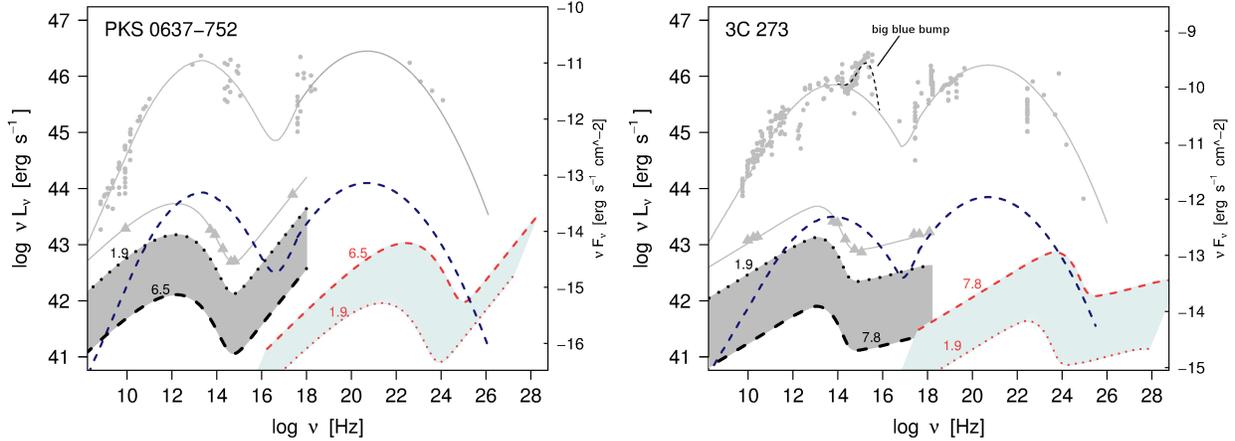}
\caption{\label{fig:tevplot}(Note: Figure description below applies to
  both panels; PKS~0637-752 is shown at left and 3C~273 is shown at
  right). The core flux points are shown as gray circles with
  phenomenological SED fit through the points drawn as a thin gray
  line. The big blue bump, visible only in 3C 373, is shown as black
  dashed line and not included in the beamed emission fits. In
  comparison, the total kpc-scale jet flux shown as gray triangles.
  The flux scale at right only applies to solid curves in the
  figure. The impression of the total dominance of the core flux is
  mainly a product of the beaming difference between the two
  components, as seen when beaming-corrected (angle-integrated)
  luminosities are plotted, rather than those assuming isotropy. Blue
  dashed line is the core fit times 1/$\delta^2$ with $\delta=15$ for
  both jets, representing the angle-integrated total core output.  The
  gray zone is the range of possible angle-integrated output for the
  knots given our current constraints on $\delta$ for the knots.
  Finally, the light blue shaded area is the range of IC/CMB emission
  from the knots possible given the same $\delta$ constraints. Note
  that the TeV emission in particular is already constrained to be
  much higher than is typical for TeV blazars ($\approx 10^{41}$ erg
  s$^{-1}$).}
\end{center}
\end{figure*}

With IC/CMB ruled out in these three cases, we must explore
alternative sources of X-ray emission for these knots. Synchrotron
self-Compton (SSC) emission was ruled out very early on as being far
too weak \citep{chartas2000}, unless very low (far from equipartition)
magnetic fields were adopted; however, the total power requirement
then becomes far too high ($>10^{49}$ erg s$^{-1}$) and the Doppler
factor very small ($<<1$) suggesting an unrealistically de-beamed jet
\citep{tav00,der04}.
A viable alternative, though requiring further exploration, is a
hadronic origin for the X-rays. Such models
\citep[e.g.][]{aharonian2002} would not be in conflict with the high
polarization measurements of \cite{cara2013}, and these models may be
tunable enough to match the X-ray flux level while avoiding the GeV
limits \citep{zhang_boettcher2013,kundu2014}.
A (leptonic) synchrotron origin for the X-rays from a second
population of electrons is also not in conflict with any of the data
in hand, and further, relaxes many of the `uncomfortable' constraints
of the IC/CMB model. Very small angles to the line of sight are not
required, and the total jet power required is considerably less
\citep{der04}, as the electron energy distribution need not be
extended to very low values. The main objection to a second
synchrotron component heretofore has simply been its unexplained
nature; \cite{schwartz2000} notes that there is no reason why a second
population of high-energy electrons should be co-spatial with the
first. However, this co-occurrence of two very different electron
populations, if the correct interpretation, is obviously a very
important clue to the particle acceleration mechanism in large-scale
jets, of which we still know little. An additional characteristic of
the second electron population, at least for the case of PKS~0637-752,
is that it requires a low-energy-cutoff in the electron energy
distribution at TeV energies, otherwise the spectrum should extend to
optical energies, contrary to the cutoff seen in the radio optical
knot emission \citep{mehta2009}. Acknowledging that further work may
be warranted in the direction of evaluating hadronic models, we focus
the rest of the paper on the implications for jet physics if the X-ray
flux in quasar jets is synchrotron emission from a separate,
high-energy electron population.


Assuming that the X-ray emission from the jets of 3C~273 and
PKS~0637-752 is synchrotron in origin, an interesting consequence
follows for our accounting of the large-scale-jet contribution to
various backgrounds, especially at TeV energies. Jet one-sidedness
clearly indicates that the kpc-scale jets are at least mildly
relativistic, and thus IC/CMB emission must occur at some level.  Due
to the very low background in the highest-energy \emph{Fermi} bands,
the flux limits reachable by \emph{Fermi}'s sky-scanning mode of
operation should allow us to eventually either detect this emission or
put very strong limits on the factor of $B/\delta$ which characterizes
the flow on the kpc scale. The current \emph{Fermi} 95\% upper limits
already constrain $\delta\lesssim7.8$ for 3C~273 and
$\delta\lesssim6.5$ PKS~0637-752, under the assumption of
equipartition magnetic fields, where we take $B\delta$ = 1.5$\times
10^{-5}$ G for PKS~0637-752 from \cite{tav00} and $B\delta$=1.0$\times
10^{-4}$ for 3C~273 from \cite{jes06}. Lower magnetic field values
would only decrease these upper limits on $\delta$. These limits are
already low enough to have interesting consequences for our
understanding of the total radiative output of AGN jets on the kpc
scale.

It is generally assumed that the radiative output of quasar jets is
dominated by that occurring at the `core', the base of the jet which
is presumed to be very near the black hole (or $\sim$ parsecs away at
most) and is therefore unresolved even in VLBI imaging. Certainly, the
observed fluxes are dominated by this part of the jet due to Doppler
beaming whenever the jet is pointed fairly along our
line-of-sight. This is depicted in Figure~\ref{fig:tevplot}, where the
core flux of both jets are shown as gray circles, and the total
kpc-scale jet as gray triangles (flux scale on right axis). The
luminosity scale at left applies to these points only under the
incorrect assumption of isotropy.  If we correct these values for
beaming, we get the real `angle-integrated' total power output from
the core and the jet, plotted as a dark blue dashed line and the gray
shaded area, respectively. VLBI observations of superluminal motions
place a lower limit on $\Gamma$=15 for the cores of both 3C~273 and
PKS~0637-752, respectively \citep{lister2013,edwards2006}.  We have
applied a correction \citep[multiplying by
  1/$\delta^2$,][]{ghisellini2010} assuming $\delta$=15 for the cores
of both jets to give the angle-integrated core luminosity (dark blue
dashed line). To calculate the angle-integrated luminosity of the
knots, we apply a lower limit value of $\delta=1.9$ which comes from
statistical arguments based on populations \citep{arshakian2004},
which gives the dotted-line upper edge to the gray shaded area, while
the current $\delta$ limits from \emph{Fermi} give the lower
dashed-line limit.  The true angle-integrated luminosity of the knots
is thus somewhere in the gray zone.

Comparing the dashed-blue curve and the gray shaded area, it is
interesting to note that the knots are apparently \emph{not}
necessarily insignificant in total output when compared to the
core. The jet power dissipated by radiative losses appears to occur
over much larger physical scales than has previously been appreciated.
Definite conclusions will require tighter constraints on the $\delta$
factors of both the core and the knots, but it is possible that
large-scale jets contribute more than the core in the UV to X-rays, in
addition to their general dominance in the radio. Large-scale jets
could thus also be a contributor to some astrophysical backgrounds. 

A further observation follows from the realization that the X-rays are
synchrotron in origin: the electrons producing the synchrotron X-rays
will themselves upscatter the CMB to produce a GeV to TeV
spectrum. The angle-integrated total power in the IC/CMB component is
shown in Figure~\ref{fig:tevplot} as a light blue shaded area.  Note
that the bounds in this case are flipped; the upper $\delta$ limit
forms the upper edge of the allowed zone. This is because IC/CMB
emission increases with the Doppler factor as $\delta^2$ even after
the angle-integrated 4$\pi$ luminosity correction is included. Even in
the minimum $\delta$=1.9 case, these jets are already constrained to
produce fluxes in excess of $10^{41}$ erg s$^{-1}$ which is the
typical angle-integrated luminosity for canonical TeV blazars
(adopting an observed luminosity of $10^{43-44}$ erg s$^{-1}$ and a
modest $\delta$=10). This is important because it has been proposed
that (lower-power) TeV BL Lacs are significant contributors to the
heating of the intergalactic medium (IGM;
e.g. \citealt{broderick2012,chang2012,lamberts2015}, but see also
\citealt{sironi2014}), where the high-energy gamma-rays ($>$ 100 GeV)
produced by these jets pair-produce off the extragalactic background
light, eventually depositing kinetic energy through plasma
instabilities. Clearly, if this mechanism is important, the
large-scale jets of powerful quasars may be a far more important
source of $>$ 100 GeV photons than TeV BL Lacs, and possibly the
dominant class of sources for blazar heating.

We have not applied an EBL correction to these spectra merely to
illustrate the total intrinsic output. However, EBL absorption is very
important at TeV energies, and would make direct observation of this
TeV component difficult. Even assuming the most optimistic case
of $\delta$=7.8 for 3C~273, at redshift 0.158 the EBL absorption
\citep{finke2010_ebl} is already high enough that it would take at
least 100 hours of observations by the future CTA mission to detect
the beamed IC/CMB component at TeV energies.  Thus it is unlikely that
many anomalous X-ray jets will have a synchrotron origin for the
X-rays directly confirmed via TeV observations, though if \emph{Fermi}
begins detecting the IC/CMB component, an upturn at the highest
energies might be visible in a few cases. The remaining best direct
observation is via polarization -- either in the UV, for those that
show the second component emerging there, or with future X-ray
polarimeters. Finally, we note that as long as \emph{Fermi} continues
to operate, the low background at the highest energies should allow
continually improving constraints on the $\delta$ factors of
large-scale jets.



\section{Conclusions}
We have shown that the expected gamma-ray emission required if quasar
jets produce X-rays via IC/CMB has been ruled out in two quasar jets,
3C~273 and PKS~0637-752, at a high significance level using
\emph{Fermi} upper limits. Examining the evidence, we favor a
synchrotron origin for the X-rays of their large-scale jets. This
considerably relaxes the power requirements for the jet away from the
near and super-Eddington values. We have shown that the limits on the
IC/CMB gamma-rays constrain the jets to have $\delta$ values of $<$7.8
for 3C~273 and $<$6.5 for PKS~0637-752 assuming equipartition. A very
interesting outcome of the synchrotron explanation of the X-rays is
the prediction that quasar jets may radiate far more at TeV energies
than TeV BL Lacs. Future work remains to see if 3C~273 and
PKS~0637-752 are outliers, or if IC/CMB can be ruled out as the source
of the anomalous X-rays for more quasar jets.

\bibliography{pks0637}

\acknowledgments
ETM acknowledges Fermi Grant NNX13AO88G.



{\it Facilities:} \facility{Fermi}, \facility{HST (ACS, NICMOS, WFPC2)}, \facility{CXO (ASIS), \facility{Spitzer}, \facility{ATCA}}.






\end{document}